\documentclass[aps,prc,twocolumn,showpacs,groupedaddress,floatfix]{revtex4}

\usepackage{graphicx}
\usepackage{dcolumn}
\usepackage{amsfonts}

\begin{document}

\title{Compressibility and equation of state of finite nuclei}

\author{A.S. Umar and V.E. Oberacker}
\affiliation{Department of Physics and Astronomy, Vanderbilt University,
             Nashville, Tennessee 37235, USA}

\date{\today}


\begin{abstract}
We present a new approach for calculating the nuclear equation of state
and compressibility for finite nuclei using the density-constrained Hartree-Fock method.
\end{abstract}
\pacs{21.60.-n,21.60.Jz}
\maketitle

\section{\label{sec:intro}Introduction}
The study of the nuclear equation of state (EOS) and the behavior
of nuclear matter under extreme conditions is crucial to our understanding
of many nuclear and astrophysical phenomena.
With the increasing availability of radioactive ion-beams~\cite{DOE02} the study of
structure and reactions of exotic nuclei are now possible, thus providing information
concerning the isospin dependence of asymmetric nuclear matter~\cite{steiner05,baran05}.
Recently, the study of the symmetry energy for such systems has been an active area of
interest~\cite{steiner05,brown98,diep03,WRR06}.

Most studies of the EOS involve infinite or semi-infinite nuclear matter and examine
the dependence of the EOS on the parametrizations of the effective interaction as well
as its relation to the macroscopic and macroscopic-microscopic models of nuclear
matter~\cite{JPB95}.
For finite nuclei the EOS near the equilibrium density can be investigated
via collective observables such as the isoscalar monopole vibrations (breathing mode)~\cite{JPB80}.
In addition, there are empirical methods
relating the compressibility of finite nuclei to that of infinite nuclear matter~\cite{MS95}.
However, the behavior of the EOS for finite nuclei far from the equilibrium density is poorly known.

In this manuscript we introduce a new method for calculating the zero temperature EOS and related
quantities for finite nuclei within the mean-field description of nuclear properties.
In Section~\ref{sec:theory} we outline the general formalism for our calculations.
Section~\ref{sec:results} discusses the application of the formalism to a sample set
of nuclei and the results obtained. The paper is concluded with a summary in Section~\ref{sec:summary}.

\section{\label{sec:theory}Formalism}

In order to study finite nuclei away from their ground state equilibrium we
take advantage of the density constrained Hartree-Fock (DCHF) method~\cite{CR85,US85}.
The {\it density constraint} is a novel numerical method that was developed in
the mid 1980's and was used to provide a microscopic
description of the formation of shape resonances in light systems~\cite{US85}.
Recently, we have used the same method to calculate heavy-ion interaction
potentials from the TDHF time-evolution of nuclear collisions~\cite{UO06a}.
In the traditional constrained Hartree-Fock (CHF) notation, density constraint
corresponds to the replacement
\begin{equation}
\lambda\hat Q \longrightarrow \lambda\hat \rho\;.
\end{equation}
The numerical procedure for implementing this constraint and the method for
steering the solution towards a specified $\rho_{0}(\mathbf{r})$ is discussed in Refs.~\cite{CR85,US85}.
The convergence property is as good if not better than the traditional CHF
calculations with a constraint on a single collective degree of freedom.

\begin{figure}[!hbt]
\begin{center}
\includegraphics*[scale=0.40]{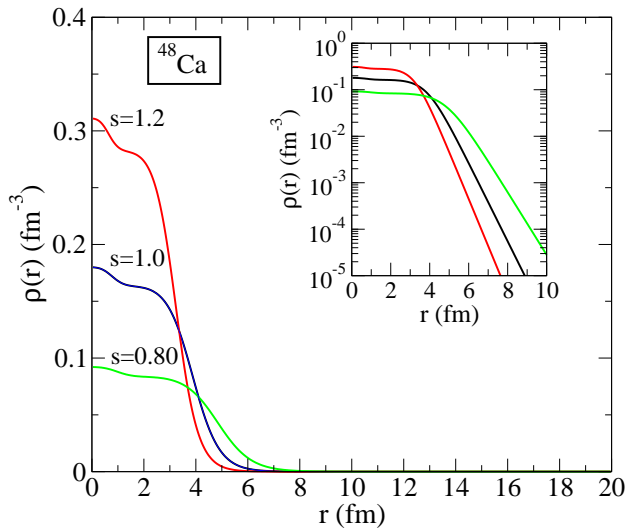}
\caption{\label{fig:density} (Color online) Scaled density obtained for the $^{48}$Ca nucleus for various
values of the scaling parameter $s$. The inset shows the same quantity on a logarithmic
scale. The force used was SLy4 without the Coulomb contribution.}
\end{center}
\end{figure}

In practice, we have obtained accurate densities for spherical nuclei using a radial
Hartree-Fock program. This density was then fitted to a parametric function of the form
\begin{equation}
\rho(r)=\frac{a_0(1+a_1r)}{1+e^{(r-a_2)/a_3}}+\frac{a_4}{1+e^{(r-a_5)/a_6}}\;,
\end{equation}
where $a_0,\ldots,a_6$ denote the parameters to be fitted to reproduce a particular density profile.
In all cases the resulting non-linear fits were indistinguishable from the fitted density.
Subsequently, the constraining density was obtained via a scale transformation of the above
density profile
\begin{equation}
\rho(r)\longrightarrow \rho(sr) \;,
\end{equation}
followed by a renormalization to produce the correct mass number for the nucleus under
study. This scaling results in the compression of the bulk and stretching of the
surface~\cite{JPB80}. Naturally, the minimum of all EOS curves occur at the
unconstrained Hartree-Fock minimum corresponding to $s=1$.
The value of the scale $s$ was generally limited to the range $(0.8,1.2)$.
In Fig.~\ref{fig:density} we show the change in the density
for various values of the scaling parameter $s$ in the case of the $^{48}$Ca nucleus.
As the primary interaction we have used the Skyrme SLy4 force~\cite{CB98}, with and
without the Coulomb term, and including all of the spin-dependent terms.
We have performed the calculations using our new Hartree-Fock program
discussed in Ref.~\cite{UO06b}.

In dealing with a finite nucleus we have faced a conceptual problem of deciding what density
value to use for plotting the density dependence of the EOS. Unless otherwise stated we have
used the {\it central density} as the reference density value for each value of the
scaling parameter $s$. Alternatively, one can choose the nuclear matter equilibrium
density, 0.16~fm$^{-3}$, as the reference density from the density profile.
The calculated values for incompressibility is slightly dependent on the choice of
the reference density point due to the structure in the density profile. Our calculations
show that this is about 10\% or less and becomes negligible for heavy systems.
One disadvantage of using the central density as the reference value is that the equilibrium
densities for different nuclei do not occur at the same value, thus making the calculation of
the symmetry energy, which is essentially taking the difference of two EOS curves along an
isotope chain, erroneous. For this reason, in the calculation of the symmetry energy we
have used the density value 0.16~fm$^{-3}$ as the equilibrium density.

In order to extract the incompressibility coefficient, $K_{\mathrm{A}}$, we have expanded
the EOS (binding energy per particle as a function of density) around the equilibrium
density $\rho_0$ using the expression
\begin{equation}
\frac{E(\rho)}{A}=\frac{E_0}{A}+\frac{K_{\mathrm{A}}}{18\rho_0^2}(\rho-\rho_0)^2+\ldots\;,
\end{equation}
where $E_0$ is the binding energy at the equilibrium density, and $K_{\mathrm{A}}$ is the
incompressibility coefficient
\begin{equation}
K_{\mathrm{A}}=\left. 9\rho_0^2\frac{\partial^2(E/A)}{\partial\rho^2}\right|_{\rho=\rho_0}\;.
\end{equation}

We found in practice that this expression provides an excellent fit except at the extreme
values of the density. To fit the entire curve perfectly a small linear contribution
as well as a cubic term could be included. The results for the incompressibility is
approximately 5\%-10\% higher if the full curve is used in the fit.
Finally, to extract the symmetry energy
we use the expression
\begin{equation}
{\cal E}_{\mathrm{sym}}(\rho) = \frac{E(\rho,\alpha)}{A} - \frac{E(\rho,0)}{A}\;,
\label{eq:symmetry1}
\end{equation}
where the isospin asymmetry parameter is defined as $\alpha=(N-Z)/A$.
The $\alpha$ dependence of the symmetry energy is generally acknowledged to be
\begin{equation}
{\cal E}_{\mathrm{sym}}(\rho) = S(\rho)\alpha^2 + O(\alpha^4)\;,
\label{eq:symmetry2}
\end{equation}
where the higher order terms in $\alpha$ are assumed to be small.
Traditionally, the symmetry energy can also be expanded around the equilibrium
density as
\begin{displaymath}
{\cal E}_{\mathrm{sym}}(\rho) = {\cal E}_{\mathrm{sym}}(\rho_0) + \frac{L}{3\rho_0}(\rho-\rho_0)+\frac{K_{\mathrm{sym}}}{18\rho_0^2}(\rho-\rho_0)^2+\ldots
\end{displaymath}
where we have defined the quantities $L$ and $K_{\mathrm{sym}}$, which
are related to the symmetry pressure and symmetry compressibility
\begin{displaymath}
L=3\rho_0\left( \frac{\partial{\cal E}_{\mathrm{sym}}} {\partial\rho}\right)_{\rho=\rho_0}\;,\;
K_{\mathrm{sym}}=9\rho_0^2\left(\frac{\partial^2{\cal E}_{\mathrm{sym}}} {\partial\rho^2}\right)_{\rho=\rho_0}\;.
\end{displaymath}

\section{\label{sec:results}Numerical studies}
Using our approach we have first investigated the EOS for the $^{16}$O nucleus,
with two forces, SLy4 and SkM$^*$~\cite{BQ82}, without the Coulomb interaction.
We found very little difference between the two forces as expected since most
modern Skyrme forces have similar nuclear matter incompressibility values. For
the incompressibility we find $K_{\mathrm{A}}=116$~MeV.
\begin{figure}[!hbt]
\begin{center}
\includegraphics*[scale=0.40]{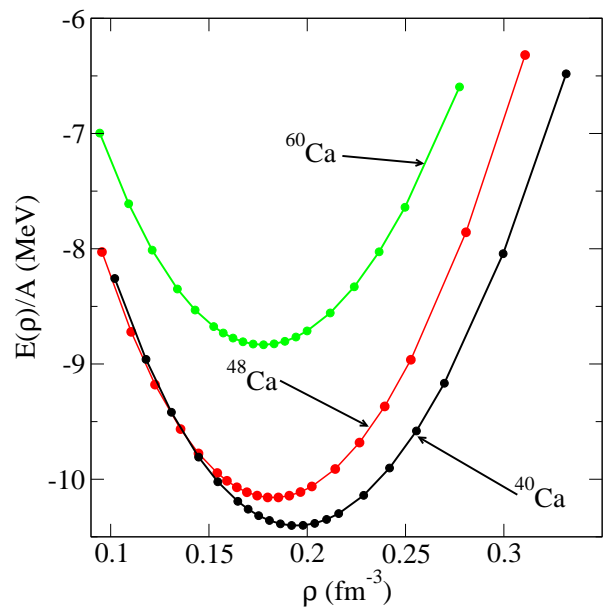}
\caption{\label{fig:Ca1} (Color online) EOS for the $^{40}$Ca, $^{48}$Ca, and $^{60}$Ca nuclei as a
function of density. Please see the comments in the manuscript regarding the density
values. The force used was SLy4 without the Coulomb contribution.}
\end{center}
\end{figure}
We have then repeated the
same two force study for the $^{40}$Ca nucleus, yielding an incompressibility of
$K_{\mathrm{A}}=141$~MeV. In order to investigate the behavior of neutron
rich systems we have performed calculations for $^{48}$Ca and $^{60}$Ca systems.
In Fig.~\ref{fig:Ca1} we show the EOS for all of these Ca nuclei without the Coulomb force.
Again, we stress that the density scale shown in Fig.~\ref{fig:Ca1} is determined
by choosing the central density in the density profile of each
nucleus. Consequently, the equilibrium value for the EOS is different for each nucleus.
As we can see from Fig.~\ref{fig:Ca1}, the $^{40}$Ca and $^{48}$Ca systems have a very similar
EOS behavior, the primary difference being the energy shift due to the difference in
the two binding energies. The calculated incompressibility of $^{48}$Ca is
$K_{\mathrm{A}}=155$~MeV, only slightly higher than the one for $^{40}$Ca. However,
the situation for $^{60}$Ca is significantly different, in addition to the large
shift in the energy scale the incompressibility decreases to a value of
$K_{\mathrm{A}}=136.5$~MeV, indicating a somewhat softer nucleus. The values for the
incompressibility obtained here, perhaps with the exception of the $^{60}$Ca
system, are in general agreement with those given in Ref.~\cite{MS95}.

We have repeated some of the calculations by including the Coulomb interaction
as well. In general, EOS curves are shifted up due to the decrease in the binding
energy per nucleon. In addition, we see a small decrease in the incompressibility
modulus. For $^{40}$Ca we find $K_{\mathrm{A}}=138.8$~MeV. The incompressibility
for $^{48}$Ca decreases to $K_{\mathrm{A}}=147.7$~MeV, and this value drops down
to $K_{\mathrm{A}}=123.5$~MeV for $^{60}$Ca.
\begin{figure}[!hbt]
\begin{center}
\includegraphics*[scale=0.35]{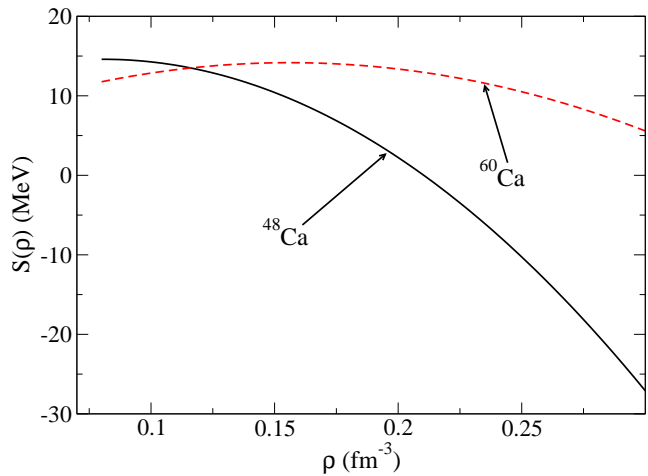}
\caption{\label{fig:Sym1} (Color online) The quantity \protect$S(\rho)$ for the $^{48}$Ca and $^{60}$Ca nuclei as a
function of density.}
\end{center}
\end{figure}

We have also investigated the symmetry energy obtained from Eq.~\ref{eq:symmetry1}, which is
essentially the difference between the curves shown in Fig.~\ref{fig:Ca1}. The results
for the quantity $S(\rho)$ defined in Eq.~\ref{eq:symmetry2},
corresponding to nuclei $^{48}$Ca and $^{60}$Ca having $\alpha$ values of 0.16667 and
0.33334, respectively, are shown in Fig.~\ref{fig:Sym1}. As we observe again, the two
systems behave very differently. Unlike infinite nuclear matter the curves do not cross
at the equilibrium density since the binding energy per nucleon is different for
different nuclei.
The values of $S$ around the equilibrium density are about 10-14~MeV, which is considerably
lower than the estimated range for nuclear matter value of 32~MeV for the SLy4 force.
Using the results of Fig.~\ref{fig:Sym1} we have extracted the quantities $L$ and
$K_{\mathrm{sym}}$ for $^{48}$Ca and $^{60}$Ca. The results show a very large variation
between the two systems. For $^{48}$Ca we find 66.6~MeV and -403~MeV for $L$ and
$K_{\mathrm{sym}}$, while for $^{60}$Ca these numbers become 1.55~MeV and -193~MeV,
respectively.

\section{\label{sec:summary} Conclusions}
We have introduced a new method for calculating the nuclear EOS for finite nuclei
including all of the terms in the nuclear effective interaction. The calculated
values agree well with known values obtained by other means, such as those deduced
from experimental giant monopole resonances. Much work has gone into understanding
the effects of various components of the effective interaction which vanish or become
very small in the infinite nuclear matter limit. With the increased availability of
new radioactive neutron and proton rich nuclei the study of EOS and symmetry energy
along isotope chains of finite nuclei have become more urgent. We believe that the
density constrained HF method is a step in this direction.

\begin{acknowledgments}
This work has been supported by the U.S. Department of Energy under grant No.
DE-FG02-96ER40963 with Vanderbilt University.
\end{acknowledgments}

\end{document}